\documentclass[aps,showpacs,nofootinbib,superscriptaddress]{revtex4}

\usepackage{graphicx}
\usepackage{graphics}
\usepackage{amssymb}
\usepackage{amsmath}

\newcommand{\de}{d}

\newcommand{\eps}{\epsilon}

\newcommand{\g}{\gamma}
\newcommand{\sig}{\sigma}
\newcommand{\tr}{\mbox{Tr}\,}
\newcommand{\ii}{i}

\newcommand{\pslash}{\rlap{/} p}

\newcommand{\st}{{ T}}
\newcommand{\sa}{{ A}}

\begin{document}

\title{
Reviewing model calculations of the Collins fragmentation function
}

\author{Daniela Amrath}
\email{daniela.amrath@desy.de}
\affiliation{Institut f{\"u}r Theoretische Physik II, Ruhr-Universit{\"a}t Bochum,
D-44780 Bochum, Germany}
\affiliation{
Deutsches Elektronen-Synchroton DESY, D-22603 Hamburg, Germany}

\author{Alessandro Bacchetta}
\email{alessandro.bacchetta@physik.uni-regensburg.de}
\affiliation{Institut f{\"u}r Theoretische Physik, Universit{\"a}t Regensburg,
D-93040 Regensburg, Germany}

\author{Andreas Metz}
\email{andreas.metz@tp2.ruhr-uni-bochum.de}
\affiliation{Institut f{\"u}r Theoretische Physik II, Ruhr-Universit{\"a}t Bochum,
D-44780 Bochum, Germany}

\begin{abstract}
The Collins fragmentation function describes a left/right asymmetry in 
the fragmentation of a
transversely polarized quark into a hadron in a jet. 
Four different model calculations of the Collins function have been presented
in the literature.
While based on the same concepts, they lead to different results and in
particular to
different signs for the Collins function. 
The purpose of the present
work is to review the features of these models and
correct some errors made in previous calculations. A full study of
the parameter dependence and the possible modifications to these models is
beyond the scope of the paper. However, some general
conclusions are drawn.

\end{abstract}



\date{\today}

\maketitle

\section{Introduction}
\label{s:intro}

The Collins fragmentation function~\cite{Collins:1993kk} contributes to
several single-spin asymmetries in hard scattering 
reactions, e.g.\ semi-inclusive deep inelastic scattering
(DIS)~\cite{Mulders:1995dh,Boer:1997nt}, 
proton-proton collisions~\cite{Anselmino:1994tv}
and electron-positron annihilation into hadrons~\cite{Boer:1997qn}. Several phenomenological works on the Collins function have been
published in the past
years (see,e.g.,
Refs.~\cite{Efremov:1998fu,Efremov:2001cz,Ma:2002ns,Efremov:2002ut,Anselmino:2004ky}), based 
on experimental data coming from the above-mentioned
processes~\cite{Airapetian:2000tv,Airapetian:2001eg,Avakian:2003pk,Adams:1991cs,Adams:2003fx}.
In particular, 
the HERMES collaboration has recently extracted a
component of the transverse spin asymmetry in semi-inclusive DIS~\cite{Airapetian:2004tw}, providing up to now the cleanest evidence of the existence of a
nonzero Collins function. Although the interpretation of the data is still 
under debate -- because of the unexpected relative behavior of $\pi^+$, $\pi^0$
and $\pi^-$ asymmetries --  the favored Collins function
for pions seems to be positive and small~\cite{Efremov:2004hz}. The COMPASS
collaboration performed the same kind of measurement using a deuteron target,
obtaining asymmetries consistent with zero~\cite{Alexakhin:2005iw}. 
This could be due simply to the difference between the targets. 

Four model calculations of the Collins function for the fragmentation of a
quark into a pion have been presented so
far in the
literature~\cite{Bacchetta:2001di,Bacchetta:2002tk,Gamberg:2003eg,Bacchetta:2003xn}
and used to make predictions and/or compare to available data~\cite{Schweitzer:2003yr,Gamberg:2003eg,Gamberg:2003pz,Gamberg:2004wt}. 
All of them produce the necessary imaginary parts by adopting a 
simple model for the fragmentation 
process at tree level 
and inserting one-loop corrections. Two possibilities have been
investigated for the tree-level amplitudes and two possible kinds of one-loop
corrections, for a total of four different models: pseudoscalar
pion-quark coupling with pion loops~\cite{Bacchetta:2001di} and with
gluon loops~\cite{Gamberg:2003eg}; pseudovector pion-quark coupling with
pion loops~\cite{Bacchetta:2002tk} and with gluon
loops~\cite{Bacchetta:2003xn}. We discovered an overall sign error in
the Collins function calculated in 
Refs.~\cite{Bacchetta:2001di,Bacchetta:2002tk,Bacchetta:2003xn},
and a more fundamental error in
Ref.~\cite{Gamberg:2003eg}, as we are going to discuss in the paper.

The calculation in Ref.~\cite{Gamberg:2003eg} makes also use of Gaussian 
form factors at the pion-quark vertex. In the present analysis, we
refrain ourselves from taking into account different types of form factors and
we limit ourselves to showing the features of the models when using point-like
vertices. It would be interesting to study how our results change upon the
introduction of form factors of different types, which is however beyond the
scope of the present article. Moreover, such models are purely
phenomenological and cannot be derived from a Lagrangian of a microscopic
model. 

The paper is organized as follows: in Sec.~\ref{s:ps} we use a 
pseudoscalar pion-quark coupling to calculate the unpolarized fragmentation
function $D_1$, as well as the Collins fragmentation function generated by 
pion and
gluon loops. In Sec.~\ref{s:psvec} we repeat the same calculations using a 
pseudovector pion-quark coupling. In Sec.~\ref{s:asymm} we discuss numerical
results obtained with both versions of the pion-quark coupling and, in
particular, we give some estimates of the Collins 
single transverse spin asymmetry in semi-inclusive DIS. 

\section{Pseudoscalar pion-quark coupling}
\label{s:ps}

Fragmentation functions can be calculated from the correlation function
$\Delta(z,\vec{k}_\st)$~\cite{Boer:2003cm}, 
\begin{equation} \begin{split} 
\Delta(z,\vec{k}_\st) &= \frac{1}{4z} \int \de k^+ \;\Delta
(k,p)\,\bigg|_{k^-=p^-/z} \\
         & =\sum_X \, \int
        \frac{\de\xi^+ \de^2\vec{\xi}_\st}{4z (2\pi)^{3}}\; e^{+\ii k \cdot \xi}
       \langle 0|
{\cal U}^T{[\infty_T, \vec{\xi}_T;-\infty^+]}\;
{\cal U}^+{[-\infty^+, \xi^+;\vec{\xi}_T]}
\,\psi(\xi)|\pi, X\rangle 
\\
&\qquad\times
\langle \pi, X|
             \bar{\psi}(0)\,{\cal U}^+{[0^+, -\infty^+;0_T]}\;
{\cal U}^T{[0_T, \infty_T; -\infty^+]}
|0\rangle \bigg|_{\xi^-=0}\,.    
\label{e:delta}
\end{split} \end{equation}  
The notation ${\cal U}^+{[a^+,b^+;c_T]}$ indicates a gauge link running along
the plus direction from $(0^-,a^+,c_T)$ to $(0^-,b^+,c_T)$, while ${\cal
  U}^T{[a_T,b_T;c^+]}$ indicates a gauge link running along
the transverse direction from $(0^-,c^+,a_T)$ to $(0^-,c^+,b_T)$. The
definition written above applies to the correlation function appearing in
semi-inclusive DIS, while in $e^+e^-$ annihilation all occurrences of
$-\infty^+$ in the gauge links should be replaced by $\infty^+$. However,
in Ref.~\cite{Collins:2004nx} it was shown that by means of a certain contour
deformation one can derive factorization in such a way that both the
fragmentation functions in semi-inclusive DIS and in $e^+e^-$ annihilation
have future-pointing gauge links. This universality of fragmentation functions
was also observed earlier in the context of a specific model
calculation~\cite{Metz:2002iz}.

In the rest of the article we shall utilize the Feynman gauge, in which the
transverse gauge links ${\cal U}^T$ give no contribution and can be
neglected~\cite{Ji:2002aa,Belitsky:2002sm}. 

\begin{figure}
        \centering
        \includegraphics[width=3.5cm]{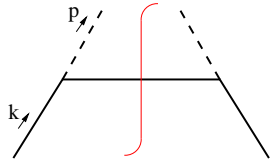}
        \caption{Tree-level cut diagram describing the fragmentation 
                of a quark into a pion. This diagram is common to all models,
                but the specific form of the pion-quark vertex can change.} 
        \label{f:born}
        \end{figure}

The tree-level diagram describing the fragmentation of a quark into a pion 
is depicted in Fig.~\ref{f:born}. In the first part of this work, the pion-quark
vertex is taken to be $g \gamma_5 \tau_i$, where $\tau_i$ are the generators
of the SU(2) flavor group.\footnote{Note that in
  Ref.~\cite{Bacchetta:2001di} the isospin structure was neglected, since it
  was not relevant to the purpose of that paper. This
  leads to different overall numerical factors in some of the final results.}
We assume the coupling to be point-like. This assumption is of course not
appropriate at large transverse momenta of the pion. In fact, when integrating
the fragmentation functions over $k_T$ divergences occur. Therefore, we
impose  a
cutoff on the virtuality of the incoming quark, and study the dependence on
the cutoff in some detail. 
A different approach would be to insert
form factors. This could sensibly change the behavior of the fragmentation
functions compared to our results.

\subsection{Unpolarized fragmentation function}
\label{s:d1ps}

We briefly reproduce the results already obtained in
Ref.~\cite{Bacchetta:2001di}, but we present also a discussion of the
parameter dependence of our results. 
Here and in the next sections, 
all results are for, e.g., the transition $u \to \pi^0$.
An additional isospin factor of 2 has to be included for, e.g., the 
transition $u \to \pi^+$.
The definition of the unpolarized fragmentation function is
\begin{equation} 
D_1(z,z^2 \vec{k}^2_{\st}) = 
              \tr[ \Delta (z,\vec{k}_\st)\, \g^-] \,.
\label{e:d1} 
\end{equation}
We compute the unpolarized fragmentation functions at tree level only, i.e.\
only using the diagram of Fig.~\ref{f:born}. This is not entirely consistent
with the fact that one-loop corrections need to be introduced in order to
calculate the Collins function. We believe that the corrections to our final
results will be small, though it would be appropriate to check in which
kinematical region this statement holds. 
The result obtained from the calculation of the tree-level diagram is 
\begin{equation} 
D_1(z,z^2 \vec{k}^2_{\st}) = \frac{1}{z}\frac{g^2}{16 \pi^3}\,
        \frac{\vec{k}_{\st}^2 +m^2}{(\vec{k}_{\st}^2 +m^2+\frac{1-z}{z^2}m_{\pi}^2)^2} \,.
\label{e:unpolfrag}
\end{equation}

The integrated unpolarized fragmentation function $D_1(z)$ 
is defined as
\begin{equation} 
D_1(z)= \pi \int_0^{\vec{K}^{2}_{\st\,{\rm max}}} \de \vec{K}^2_{\st}\; D_1(z,\vec{K}^2_{\st}),
\end{equation} 
where $\vec{K}_{\st}=- z \vec{k}_{\st}$ denotes the transverse momentum of the 
outgoing hadron with respect to the quark direction. 
The upper limit on the $\vec{K}^2_{\st}$ integration is set by the cutoff on
the fragmenting quark virtuality, $\mu^2$, and corresponds to
\begin{equation} 
\vec{K}^{2}_{\st\,{\rm max}}=z \, (1-z)\,\mu^2 -z\,m^2-(1-z)\,m_\pi^2 \,.
\end{equation} 
The analytic result for the integrated fragmentation function is
\begin{align} 
D_1(z)&=\frac{g^2}{16\, \pi^2}\Biggl[
z\,\ln \left(\frac{(1-z)\,(\mu^2  -m^2)}{z\,\left(m^2+ m_{\pi}^2\,\frac{1 - z}{z^2}\right)}
         \right)  
-
 m_{\pi}^2\,\frac{z \, (1-z)\,\mu^2 -z\,m^2-(1-z)\,m_\pi^2}
{z^2\,\left(\mu^2  -m^2 \right) \,
         \left(m^2+ m_{\pi}^2\,\frac{1 - z}{z^2} \right) } 
\Biggr].
\end{align}

\begin{figure}
        \centering
        \includegraphics[width= 6cm]{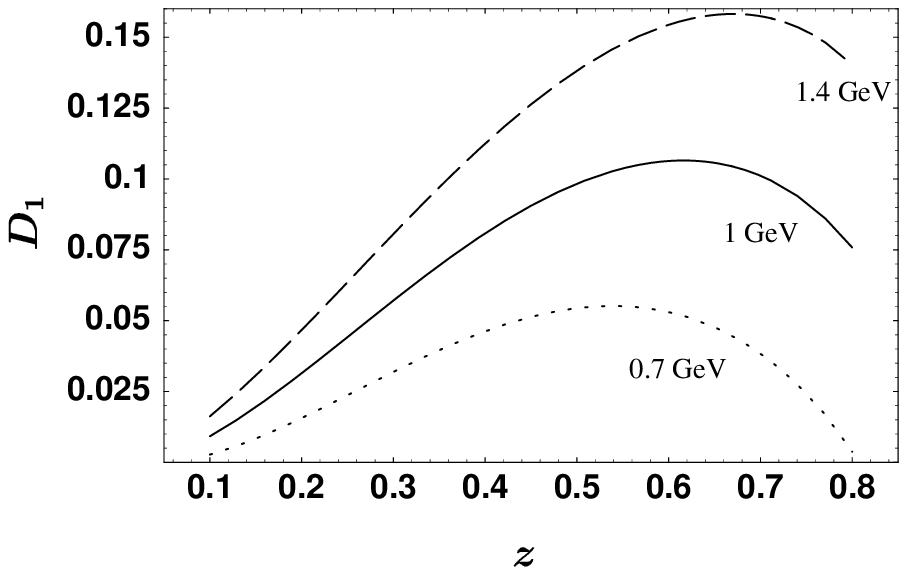}
        \hfil
        \includegraphics[width= 6cm]{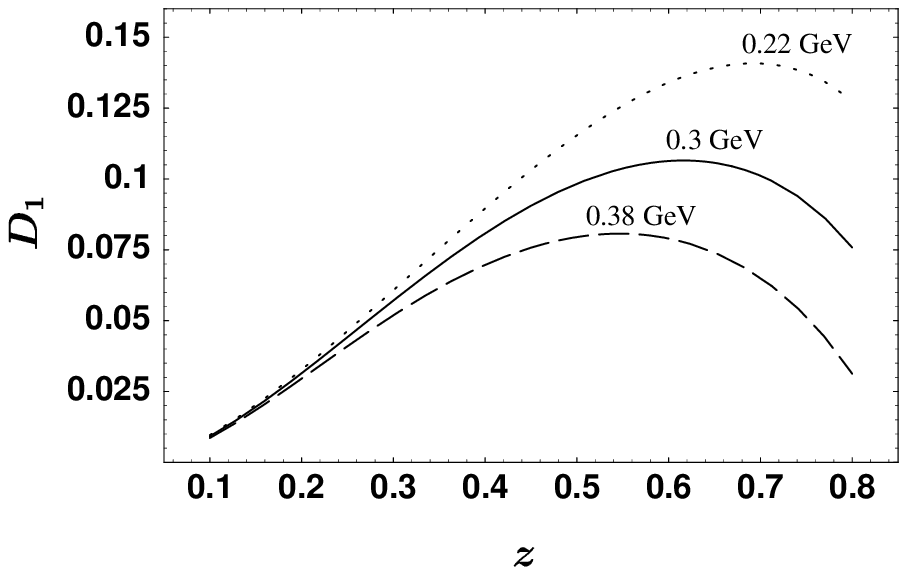}
        \caption{Unpolarized fragmentation function $D_1^{u\to \pi^+}$ in a
          fragmentation model with pseudoscalar pion-quark coupling. 
          Left panel: dependence on the parameter
          $\mu$ (for $m=0.3$ GeV). Right panel: dependence on the parameter 
          $m$ (for $\mu=1$ GeV).} 
        \label{f:D1ps}
        \end{figure}

In Fig.~\ref{f:D1ps} we show the result of the model calculation of the
function  $D_1^{u\to \pi^+}$ for a choice of the
coupling constant $g=3$ and for different values of the parameters $\mu$ and
$m$. 
Apart from the trivial dependence on the coupling strength, 
essentially an increase of the cutoff or a decrease of the quark mass 
makes the function bigger, without sensibly changing the $z$ dependence.
The shape of the unpolarized fragmentation function is very far from standard
parameterizations extracted from phenomenology (see, e.g.\
Ref.~\cite{Kretzer:2001pz}), even from a qualitative point of view. 
As mentioned before, different behaviors can be obtained by modifying the model through
the insertion of form factors, as can be seen comparing our results with those of Ref.~\cite{Gamberg:2003eg}.

\subsection{Collins function from pion loops}

We use the following definition of the Collins
function~\cite{Mulders:1995dh},  
in agreement with the ``Trento
conventions''\cite{Bacchetta:2004jz},
\begin{equation}
\frac{\eps_{\st}^{ij} k_{\st j}}{m_{\pi}}
              \, H_1^{\perp}(z,z^2 \vec{k}^2_{\st}) 
=             \tr[ \Delta(z,\vec{k}_\st)\, \ii \sig^{i-}\g_5]\,.  
\label{e:col1}
\end{equation}  

\begin{figure}
        \centering
        \includegraphics[width=9cm]{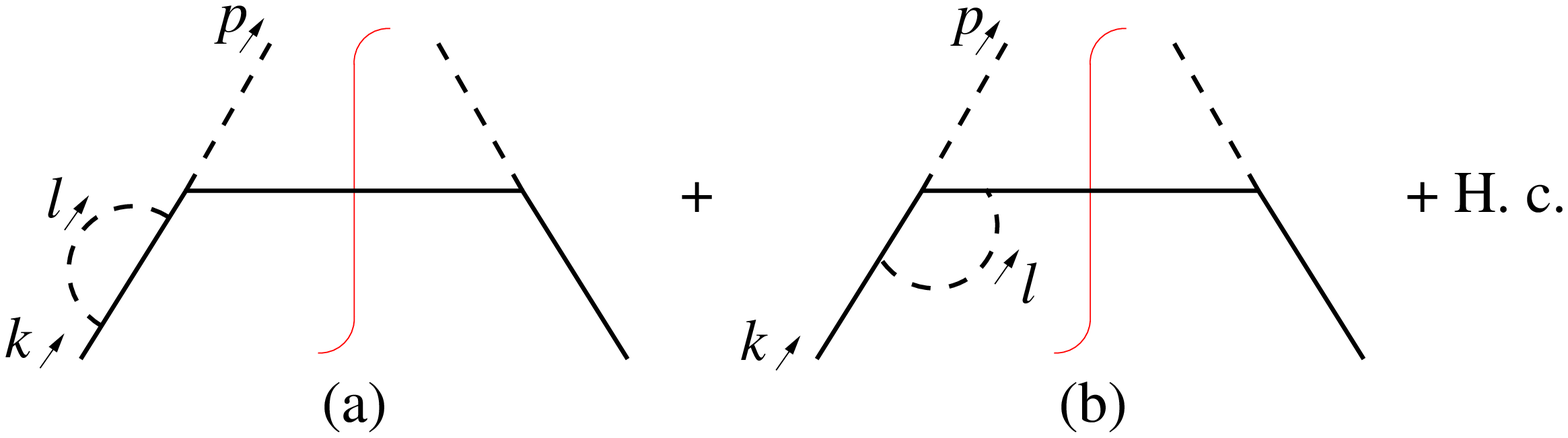}
        \caption{Single pion-loop corrections to the fragmentation of a quark
                into a pion.}
        \label{f:pspion}
        \end{figure}

The Collins function receives contributions only from the interference between
two amplitudes with different imaginary parts. In our case, the tree-level
amplitude is real and the necessary imaginary parts are generated by the
inclusion of one-loop corrections. Such corrections contain imaginary parts if
and only if it is kinematically possible 
that the particles in the loop go on-shell.   
In this section, we will make use of pion loops. Keeping in mind that the initial quark is virtual and time-like, the only two diagrams contributing to the Collins function at one-loop are the ones 
shown in Fig.~\ref{f:pspion}, plus their Hermitian conjugates. We refer to these two diagrams as self-energy (a) and vertex (b) corrections.

The explicit calculation of the one-loop diagrams leads to the following result for the Collins function:
\begin{equation}
H_1^{\perp}(z,z^2 \vec{k}_T^2) = 
 - \frac{g^2}{8 \pi^3} \frac{m_\pi}{1-z} \frac{m}{k^2 - m^2}
 \bigg({\rm Im} \, \sigma_{PS}^{\pi} 
     + {\rm Im} \, \gamma_{1,PS}^{\pi} \bigg) 
     \bigg |_{k^2 = \vec{k}_T^2 \frac{z}{1-z} + \frac{m^2}{1-z} + \frac{m_{\pi}^2}{z}},
\end{equation}
where we distinguished the contributions from diagram (a) and (b), being
respectively 
\begin{align} 
{\rm Im} \, \sigma_{PS}^{\pi} & = 
 \frac{3 g^2}{16 \pi^2}  \frac{1}{k^2 - m^2}
 \bigg( 1 - \frac{m^2 - m_{\pi}^2}{k^2} \bigg) I_{1,\pi}
, 
\label{e:sigmapspi}\\
{\rm Im} \, \gamma_{1,PS}^{\pi} & =  
 \frac{g^2}{8 \pi^2} \frac{k^2 - m^2 + m_{\pi}^2}{\lambda_{\pi}}
 \bigg( I_{1,\pi} + (k^2 - m^2 - 2 m_{\pi}^2) \, I_{2,\pi} \bigg).
\label{e:gamma1pspi}
\end{align} 
In the above formulae we have used the integrals
\begin{align} 
\begin{split}
I_{1,\pi}  & =  \int d^4l \, \delta(l^2 - m_{\pi}^2) \, 
                             \delta((k - l)^2 - m^2)
=  \frac{\pi}{2 k^2} \, \sqrt{\lambda_{\pi}} \; 
                            \theta (k^2 - (m + m_{\pi})^2)        
\end{split}
, \\
\begin{split}
I_{2,\pi} & =  \int d^4l \, \frac{\delta(l^2 - m_{\pi}^2) \, \delta((k - l)^2 - m^2)}
                                  {(k - p - l)^2 - m^2}
=  - \frac{\pi}{2 \sqrt{\lambda_{\pi}}} \,
 \ln \bigg( 1 + \frac{\lambda_{\pi}}{k^2 m^2 - (m^2 - m_{\pi}^2)^2} \bigg) \,
 \theta (k^2 - (m + m_{\pi})^2),        
\end{split}
\end{align} 
and the definition
$
\lambda_{\pi} =  \lambda(k^2,m^2,m_{\pi}^2) 
              = [k^2 - (m + m_{\pi})^2] [k^2 - (m - m_{\pi})^2]
$.

Comparing these results with the original publication~\cite{Bacchetta:2001di},
we notice that there is an overall sign error in the final result for the
Collins function. Furthermore, due to the introduction of the isospin
structure, which was neglected in 
the original work, the contribution of the 
self-energy correction, Eq.~(\ref{e:sigmapspi}), 
has an extra factor 3, while 
the contribution of the vertex correction, Eq.~(\ref{e:gamma1pspi}), has a
different sign.

\begin{figure}
        \centering
        \includegraphics[width= 6cm]{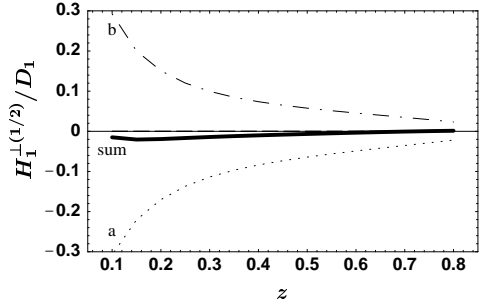}
        \caption{Contributions to $H_1^{\perp (1/2)}/D_1$ from the diagrams
          of Fig.~\ref{f:pspion} and their sum.} 
        \label{f:coll_ps_p}
        \end{figure}

In Fig.~\ref{f:coll_ps_p} we present numerical estimates of the following
quantity
\begin{equation}        
 \frac{H_1^{\perp
(1/2)}(z)}{D_1(z)} \equiv
\frac{\pi}{D_1(z)} \int \de \vec{K}_{\st}^2\, \frac{|\vec{K}_{\st}|}{2 z m_{\pi}}\,
H_1^{\perp}(z,\vec{K}_{\st}^2) \,,
\label{e:ratiomom}
\end{equation} 
separately for each diagram in Fig.~\ref{f:pspion}. The most prominent feature to be noticed is that 
the contributions of 
diagrams (a) and (b) have similar size but opposite signs, causing a strong
cancellation and giving rise
to a very small Collins function. We checked that this behavior is persistent even when changing the parameters $m$ and $\mu$. 

\subsection{Collins function from gluon loops}
\label{s:psg}

\begin{figure}
        \centering
        \includegraphics[width=10cm]{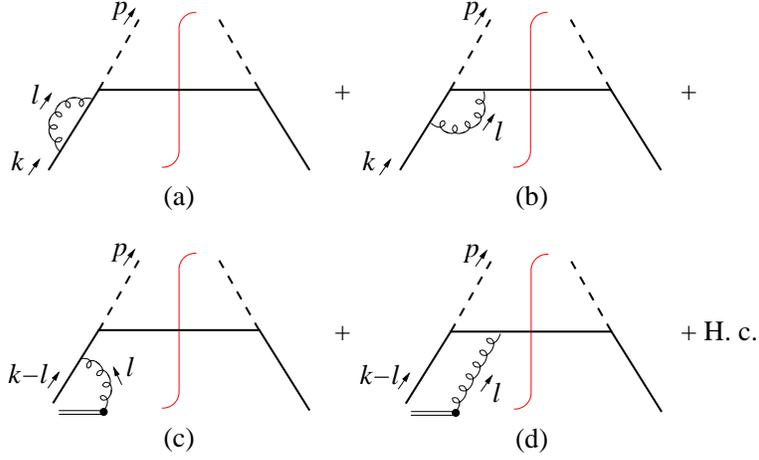}
        \caption{Single gluon-loop corrections to the fragmentation of a quark
                into a pion.}
        \label{f:psgluon}
        \end{figure}

Instead of using pion rescattering as a source of imaginary parts, it is
possible to consider gluon single-loop corrections. In fact, gluon exchange is
essential to ensure color gauge invariance of the fragmentation functions. The
diagrams involved in the calculation of the Collins function at the one-loop
level are drawn in Fig.~\ref{f:psgluon}. The self-energy (a) and vertex (b)
corrections are analogous to the previous case. The last two diagrams
represent the interaction with the quark before being struck by the photon in
semi-inclusive DIS, or with the outgoing antiquark in the case of $e^+e^-$
annihilation. We can call them the photon-vertex (c) and box diagram (d)
corrections.  At
leading order in $1/Q$, $Q$ being the virtuality of the photon, the eikonal
approximation can be applied, the quark (antiquark) can be replaced by an
eikonal line and the gluon interaction can be factorized and included in the
definition of the correlation function, giving rise precisely to the gauge
links appearing in Eq.~(\ref{e:delta}). This procedure has been discussed in
detail for distribution
functions~\cite{Ji:2002aa,Belitsky:2002sm,Boer:2002ju,Boer:2003cm}, and for fragmentation functions~\cite{Collins:2004nx}.

The Collins function obtained by the calculations of the diagrams in
Fig.~\ref{f:psgluon} can be written as
\begin{equation}
H_1^{\perp}(z,z^2 \vec{k}_T^2) = 
 - \frac{g^2}{8 \pi^3} \frac{m_\pi}{1-z} \frac{m}{k^2 - m^2}
 \bigg({\rm Im} \, \sigma_{PS}^{g} 
     + {\rm Im} \, \gamma_{1,PS}^{g} 
     + {\rm Im} \, \phi_{PS} 
     + {\rm Im} \, \xi_{PS} \bigg) 
     \bigg |_{k^2 = \vec{k}_T^2 \frac{z}{1-z} + \frac{m^2}{1-z} + \frac{m_{\pi}^2}{z}},
\end{equation}
where we distinguished the contributions from diagram (a), (b), (c), and (d), being
respectively 
\begin{align} 
{\rm Im} \, \sigma_{PS}^{g} & = 
 \frac{\alpha_s}{2 \pi} \, C_F \, 
 \frac{1}{k^2 - m^2} \bigg( 3 - \frac{m^2 
}{k^2} \bigg) I_{1,g}
, \\
 {\rm Im} \, \gamma_{1,PS}^{g} & = 
 \frac{\alpha_s}{2 \pi} \, 2 C_F \, I_{2,g} 
, \\ 
 {\rm Im} \, \phi_{PS} & =  0
 \vphantom{\frac{1}{1}}
\label{e:phiPS}
, \\ 
\begin{split} 
 {\rm Im} \, \xi_{PS} & =
 - \frac{\alpha_s}{2 \pi} \, C_F \, \frac{1}{z \vec{k}_T^2} 
 \bigg( z k^- \Big( \tilde{I}_{3,g} + (1 - z) (k^2 - m^2) \tilde{I}_{4,g}
 \Big)
- \Big( z (k^2 - m^2 + m_{\pi}^2) - 2 m_{\pi}^2 \Big) I_{2,g} \bigg). 
\end{split}
\end{align} 
These results are valid only for the case $m_g = 0$. 

Eq.~(\ref{e:phiPS}) shows that no contribution to the Collins
function arises from the photon-vertex correction of diagram (c). Even if it
is kinematically possible to have an imaginary part in this diagram, it cannot
contribute to the nontrivial Dirac structure
connected to the Collins function.

In the above formulae we have used the integrals
\begin{align}
I_{1,g} & = \int d^4l \, \delta(l^2 
) \, 
                             \delta((k - l)^2 - m^2)
= \frac{\pi}{2 k^2} \, (k^2 - m^2) \; 
                            \theta (k^2 - m^2)
 , \\
I_{2,g} & = \int d^4l \, \frac{\delta(l^2  
) \, \delta((k - l)^2 - m^2)}
                                  {(k - p - l)^2 - m^2}
= - \frac{\pi}{2 \sqrt{\lambda_{\pi}}} \,
 \ln \bigg( 1 + \frac{2 \sqrt{\lambda_{\pi}}}
 {k^2 + m^2 - m_{\pi}^2 - \sqrt{\lambda_{\pi}}} \bigg) \,
 \theta (k^2 - m^2)               
 , \\
\tilde{I}_{3,g} & = \int d^4l \, \frac{\delta(l^2  
) \, 
                           \delta((k - l)^2 - m^2)}
                           {- n \cdot l + i\varepsilon}  
\label{e:I3tilde} 
 , \\
\tilde{I}_{4,g} & = \int d^4l \, \frac{\delta(l^2  
) \, 
                           \delta((k - l)^2 - m^2)}
                           {[(k - p - l)^2 - m^2] (- n \cdot l + i\varepsilon)}.  
\label{e:I4tilde}
\end{align}
The light-like vector $n$ in the integrals $\tilde{I}_{3,g}$ and $\tilde{I}_{4,g}$
is defined via $n \cdot a = a^-$ for an arbitrary 4-vector $a^{\mu}$.
For our purpose, we need only the following linear combination of the last two
integrals
\begin{equation} 
\tilde{I}_{3,g} + (1 - z) (k^2 - m^2) \tilde{I}_{4,g} 
= \frac{\pi}{k^-} \, \ln \frac{\sqrt{k^2} (1 - z)}{m}.
\label{e:lincomb}
\end{equation}

From Eqs.~(\ref{e:I3tilde}) and (\ref{e:I4tilde}) 
it is evident that the contribution to the Collins function coming from
diagram (d) in Fig.~\ref{f:psgluon} does not come from the pole of the
eikonal propagator, but rather from the cut crossing the gluon and the incoming
quark. 
In fact, a contribution from the eikonal propagator 
would imply
a violation of universality of the fragmentation functions~\cite{Metz:2002iz}. 
This is in our opinion
a model-independent statement, merely due to kinematical conditions, in
agreement with the general statements of Ref.~\cite{Collins:2004nx}. 

At this point, we would like to comment on the model calculation presented in
Ref.~\cite{Gamberg:2003eg}. The results about diagram (d) 
obtained by the authors in that paper
are in disagreement with ours, even when taking the point-like limit of the
Gaussian form factor. We believe that this  is due to a mistake in that paper.
The Collins function obtained there,
Eq.~(11), is erroneously coming 
from the imaginary part of the eikonal propagator. This is evident
also by comparison with the results obtained for the {\em distribution} 
function $h_1^{\perp}$ in Ref.~\cite{Lu:2004hu}, Eq.~(16). Taking the 
point-like form-factor limit of the Collins function 
of Ref.~\cite{Gamberg:2003eg}, it turns out 
to be related in a simple way to the
distribution function $h_1^{\perp}$ in Ref.~\cite{Lu:2004hu}. Such a relation
should not hold, even if only diagram (d) is taken into account, since, as
already mentioned before,
the two functions receive contributions from different
cuts: the cut crossing the eikonal line and the outgoing antiquark in the
distribution function, and the cut crossing the gluon and incoming quark in the
fragmentation function.

        \begin{figure}
        \centering
        \includegraphics[width= 6cm]{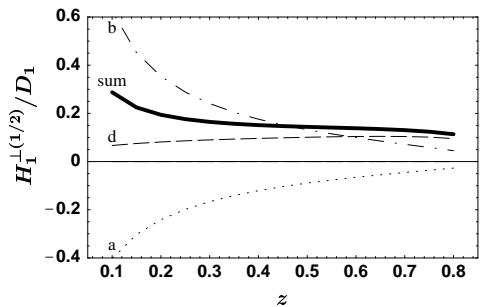}
        \caption{Contributions to $H_1^{\perp (1/2)}/D_1$ from the diagrams
          of Fig.~\ref{f:psgluon} and their sum.} 
        \label{f:coll_ps_g}
        \end{figure}

In Fig.~\ref{f:coll_ps_g}, we present numerical estimates for the quantity 
$H_1^{\perp (1/2)}/D_1$, separately for each of the diagrams of
Fig.~\ref{f:psgluon}. It is interesting to note that also in this case there is
a strong cancellation between the contribution of diagrams (a) and (b), to a large
extent independent of the parameter choice.  
At large values of $z$, the dominant part comes therefore from diagram (d), the
gauge-link box diagram. This observation may give support to the idea of calculating
the Collins function taking into account only this diagram, as
done in Ref.~\cite{Gamberg:2003eg}.

\section{Pseudovector pion-quark coupling}
\label{s:psvec}

The tree-level diagram to be used for the calculation of the unpolarized
fragmentation function is the same as before (Fig.~\ref{f:born}). However, now
the pion-quark
vertex is taken to be $g_{\sa}/(2F_{\pi}) \gamma_5 \pslash$, where $p$ is the
momentum of the outgoing pion, as drawn in the picture. When working with
on-shell particles, the pseudoscalar and pseudovector couplings are known to 
be equivalent. Here, however, we are dealing with an off-shell
fragmenting quark and the two versions of the coupling lead to completely
different results. Chiral invariance of the model is obtained by including
also a  $\pi\pi q q$ contact interaction~\cite{Manohar:1984md,Bacchetta:2002tk}. 
As we discuss below the contact term
dominates the 
 numerics of the Collins function to a large extent.
As before, we use a point-like pion-quark coupling, without form factors, 
but we impose a
cutoff in the virtuality of the incoming quark to avoid divergences.
So far, no inclusion of form factors was ever attempted on
this type of model.

\subsection{Unpolarized fragmentation function}

In this section, 
we compute the unpolarized fragmentation function $D_1$ at tree level, 
as done in
Ref.~\cite{Bacchetta:2002tk}, adding a few comments on the parameter
dependence of the outcome.
Again, all results are for, e.g., the transition $u \to \pi^0$ and an 
additional isospin factor of 2 has to be included for, e.g., the 
transition $u \to \pi^+$. The calculation of the tree-level cut diagram of
Fig.~\ref{f:born} yields the result
\begin{equation}
 D_1(z,z^2 \vec{k}^2_\st) =  
 \frac{1}{z} \frac{g_\sa^2}{4 F_\pi^2}
 \frac{1}{16\pi^3} 
\bigg( 1 - 4\frac{1-z}{z^2}
 \frac{m^2 m_\pi^2}{[\vec{k}_\st^2 +m^2 +\frac {1-z}{z^2} m_\pi^2 ]^2} \bigg). 
\end{equation}

\begin{figure}
        \centering
        \includegraphics[width= 6cm]{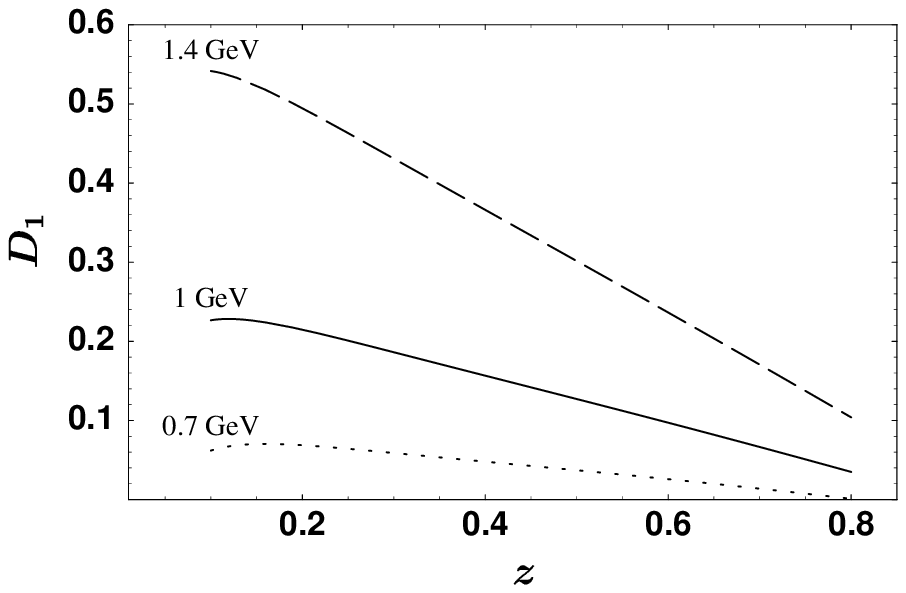}
        \hfil
        \includegraphics[width= 6cm]{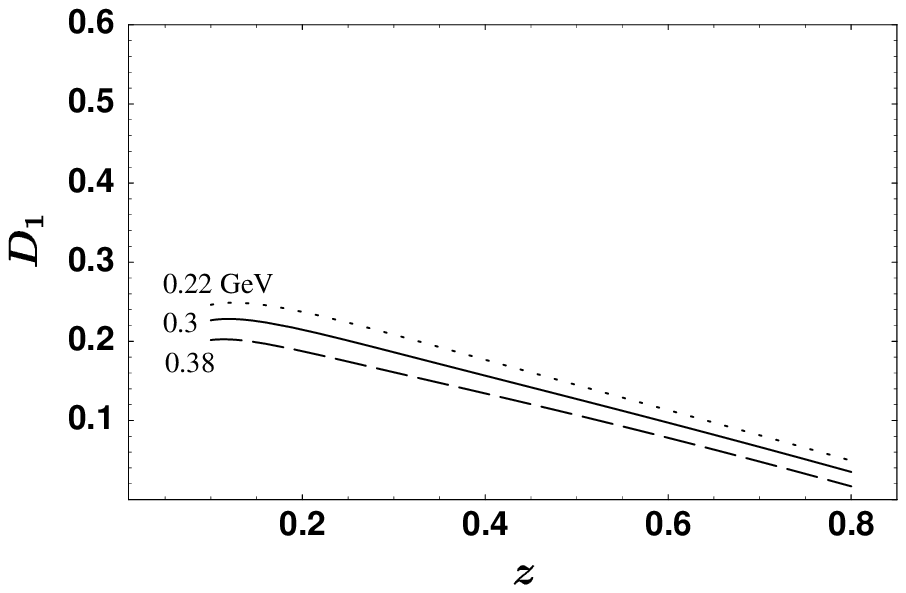}
        \caption{Unpolarized fragmentation function $D_1^{u\to \pi^+}$ in a
          fragmentation model with pseudovector pion-quark coupling. 
          Left panel: dependence on the parameter
          $\mu$ (for $m=0.3$ GeV). Right panel: dependence on the parameter
          $m$ (for $\mu=1$ GeV).} 
        \label{f:D1}
        \end{figure}

We integrate over $z \vec{k}_{\st}$ in the same manner as done in
Sec.~\ref{s:d1ps}, and obtain the integrated
fragmentation function $D_1(z)$ 
\begin{align}
D_1(z)&= \frac{g_A^2}{64\,F_{\pi}^2\,\pi^2}\,
        \left(\mu^2\,( 1 - z )  - m^2 - m_{\pi}^2\,\frac{1 - z}{z}\right)
       \,\left( 1 - \frac{4\,m^2\,m_{\pi}^2}
       {z\,\left(\mu^2 - m^2 \right) \,
         \left( m^2 + m_{\pi}^2\,\frac{1 - z }{z^2} \right)} \right). 
\end{align} 
In Fig.~\ref{f:D1} we show the result of the model calculation of the
function  $D_1^{u\to \pi^+}$ for 
$g_{\sa} =1$ and $F_\pi =93$ MeV.
We note that an increase of the value of the cutoff parameter, $\mu$,
makes the function bigger, without sensibly changing the $z$ dependence. The
dependence on the constituent quark mass, $m$, is weak. These features were
already commented on in Ref.~\cite{Bacchetta:2002tk}, where the values of
$\mu=1$  GeV and $m=0.3$ GeV were selected as best choices. This decision was
justified by comparison with a standard parameterization
of the function  $D_1^{u\to \pi^+}- D_1^{\bar{u}\to \pi^+}$ and the
experimental data on the average transverse momentum of hadrons in
fragmentation processes.

\subsection{Collins function from pion loops}

\begin{figure}
        \centering
        \includegraphics[width=14cm]{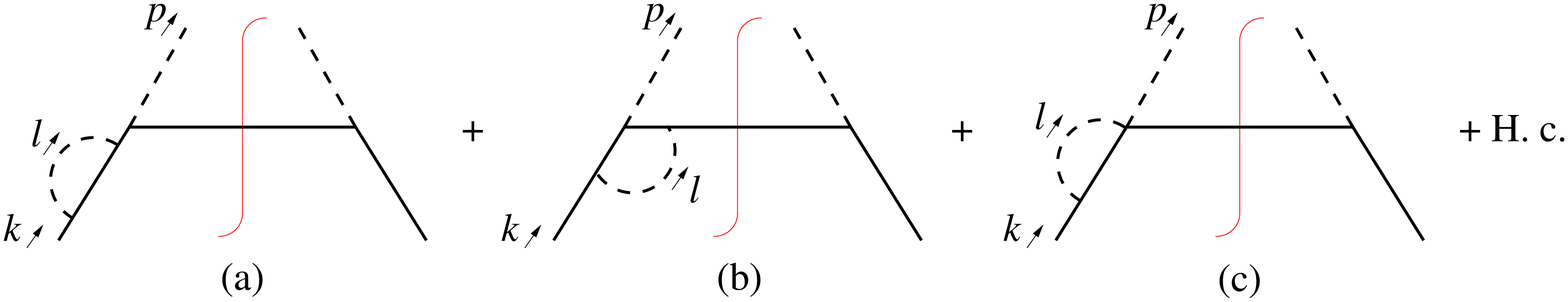}
        \caption{Single pion-loop corrections to the fragmentation of a quark
                into a pion.}
        \label{f:pvpion}
        \end{figure}

After the introduction of single
pion-loop corrections, 
all the diagrams contributing to the Collins function are
depicted in Fig.~\ref{f:pvpion}. As mentioned before, 
apart from the self-energy and
vertex corrections, chiral invariance requires the
contact-interaction term, diagram (c), which turns out to be dominant on the
others. 

%

The resulting Collins function can be written in a compact form as
\begin{equation}
H_1^{\perp}(z,z^2 \vec{k}_T^2) = 
 \frac{g_A^2}{32 \pi^3 F_{\pi}^2} \frac{m_\pi}{1-z} \frac{m}{k^2 - m^2}
 \bigg({\rm Im} \, \sigma_{PV}^{\pi} 
     + {\rm Im} \, \gamma_{1,PV}^{\pi} 
     + {\rm Im} \, \gamma_{2,PV}^{\pi} \bigg) 
     \bigg |_{k^2 = \vec{k}_T^2 \frac{z}{1-z} + \frac{m^2}{1-z} + \frac{m_{\pi}^2}{z}}.
\end{equation}
The three terms correspond respectively to the contributions of diagrams (a), (b), (c) in Fig.~\ref{f:pvpion}, and read 
\begin{align}
{\rm Im} \, \sigma_{PV}^{\pi} & =
 \frac{3 g_A^2}{32 \pi^2 F_{\pi}^2} 
 \bigg( 2 m_{\pi}^2 - \frac{1}{2} (k^2 - m^2) 
       \bigg(1 - \frac{m^2 - m_{\pi}^2}{k^2} \bigg) \bigg) I_{1,\pi}
, \\
\begin{split}
{\rm Im} \, \gamma_{1,PV}^{\pi} & = 
 \frac{g_A^2}{32 \pi^2 F_{\pi}^2} (k^2 - m^2) 
 \bigg( \frac{1}{2 k^2} (3k^2 + m^2 - m_{\pi}^2) \, I_{1,\pi}
  \\
& \quad
 + 4 m^2 \frac{k^2 - m^2 + m_{\pi}^2}{\lambda_{\pi}} 
 \Big( I_{1,\pi} + (k^2 - m^2 - 2m_{\pi}^2) \, I_{2,\pi} \Big) \bigg),  
\end{split} 
\\
{\rm Im} \, \gamma_{2,PV}^{\pi} & = 
 - \frac{2}{32 \pi^2 F_{\pi}^2} (k^2 - m^2)
    \bigg( 1 - \frac{m^2 - m_{\pi}^2}{k^2} \bigg) I_{1,\pi}.
 \end{align}
We point out that in the original publication~\cite{Bacchetta:2002tk} a sign
error was made. The sign of {\em all} results for the Collins function should 
be reversed.

\begin{figure}
        \centering
        \includegraphics[width= 6cm]{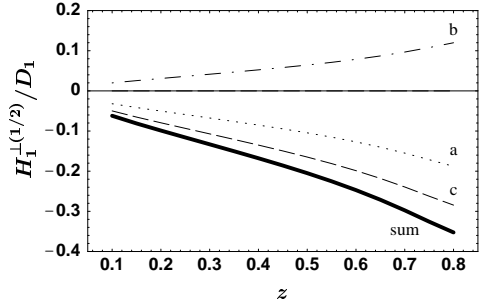}
        \caption{Contributions to $H_1^{\perp (1/2)}/D_1$
         from the diagrams
          of Fig.~\ref{f:pvpion} and their sum.} 
        \label{f:coll_pv_p}
        \end{figure}

In Fig.~\ref{f:coll_pv_p} we present numerical estimates of the
ratio $H_1^{\perp (1/2)}/D_1$,
separately for each diagram in Fig.~\ref{f:pvpion}. 
As in the previous cases, also in the present one 
the contribution from diagrams (a) and (b) (i.e.\
self-energy and vertex corrections) roughly cancel each other. The dominant
contribution to the Collins function comes therefore from diagram (c), 
i.e.\ the contact-interaction diagram. As already
mentioned before, the result of the sum of diagrams correspond to that
obtained in Ref.~\cite{Bacchetta:2002tk} (Fig.~8) {\em except} for the overall
sign.

\subsection{Collins function from gluon loops}

As discussed in Sec.~\ref{s:psg}, we can use gluon-loop corrections to
generate imaginary parts in the diagrams. The Collins function becomes
\begin{equation}
H_1^{\perp}(z,z^2 \vec{k}_T^2) = 
 \frac{g_A^2}{32 \pi^3 F_{\pi}^2} \frac{m_\pi}{1-z} \frac{m}{k^2 - m^2}
 \bigg({\rm Im} \, \sigma_{PV}^{g} 
     + {\rm Im} \, \gamma_{1,PV}^{g} 
     + {\rm Im} \, \phi_{PV} 
     + {\rm Im} \, \xi_{PV} \bigg) 
     \bigg |_{k^2 = \vec{k}_T^2 \frac{z}{1-z} + \frac{m^2}{1-z} + \frac{m_{\pi}^2}{z}},
\end{equation}
where each term represents the contribution of one of the diagrams 
in Fig.~\ref{f:psgluon}, and read 
\begin{align}
{\rm Im} \, \sigma_{PV}^{g} & =
 \frac{\alpha_s}{2 \pi} \, C_F \, 
 \bigg( 3 - \frac{m^2
}{k^2} \bigg) I_{1,g}
, \\
 {\rm Im} \, \gamma_{1,PV}^{g} & =
 - \frac{\alpha_s}{2 \pi} \, C_F \, 
 \bigg( \bigg( 1 + \frac{m^2
}{k^2} \bigg) I_{1,g}
        + 4 m_{\pi}^2 \, I_{2,g} \bigg)
, \\ 
 {\rm Im} \, \phi_{PV} & = 0
 \vphantom{\frac{1}{1}}
, \\ 
\begin{split}
 {\rm Im} \, \xi_{PV} & = 
 \frac{\alpha_s}{\pi} \, C_F \, \bigg( 2 I_{1,g} 
 + 2 k^- \Big( \tilde{I}_{3,g} + (1 - z) (k^2 - m^2) \tilde{I}_{4,g} \Big)
 + \frac{2 z m^2 - (1 - z)(k^2 - m^2)}{z^2 \vec{k}_T^2}
 \\ 
& \quad \times 
 \bigg( z k^- \Big( \tilde{I}_{3,g} + (1 - z) (k^2 - m^2) \tilde{I}_{4,g} \Big)
 - \Big( z (k^2 - m^2 + m_{\pi}^2) - 2 m_{\pi}^2 \Big) I_{2,g} \bigg) \bigg).
\end{split}
\end{align}
The above results are valid only for the case $m_g = 0$. Note the presence of
the same linear combination of  $\tilde{I}_{3,g}$ and $\tilde{I}_{4,g}$
defined in Eq.~(\ref{e:lincomb}). There is again no contribution from the
photon-vertex correction of diagram (c). Finally, 
there is no contribution to the Collins function arising from the
pole in the eikonal propagator.
We point out that in the original publication~\cite{Bacchetta:2003xn} a sign
error was made and the sing of the results for the Collins function should 
be
reversed.

\begin{figure}
        \centering
        \includegraphics[width= 6cm]{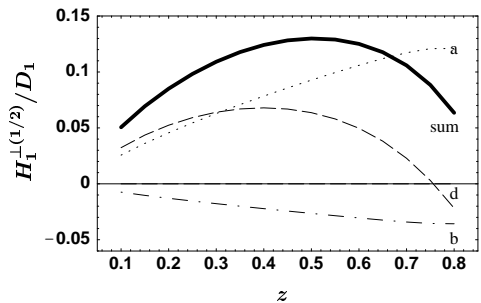}
        \caption{Contributions to $H_1^{\perp (1/2)}/D_1$  
                 from the diagrams
          of Fig.~\ref{f:psgluon} and their sum.} 
        \label{f:coll_pv_g}
        \end{figure}

Fig.~\ref{f:coll_pv_g} shows the numerical estimates for the quantity 
$H_1^{\perp (1/2)}/D_1$, separately for each of the diagrams of
Fig.~\ref{f:psgluon}. This case is different from the previous ones, 
as no strong cancellation between the different
diagrams takes place. The contribution from the vertex correction is opposite 
but much
smaller than that from the self-energy correction, a feature persisting also
for different values of the parameters. 
Hence, all contributions turn out
to be important.
The result of the sum of diagrams correspond to that
obtained in Ref.~\cite{Bacchetta:2003xn} (Fig.~4) {\em except} for the overall
sign.
This completes our review of the different models to calculate the Collins
function and the analysis of the contributions from each separate diagram. We
turn now to some numerical results.

\section{Numerical results and asymmetries}
\label{s:asymm}

In this section, we present numerical results for the quantity $H_1^{\perp
  (1/2)}/D_1$, for different values of the relevant parameters $m$ and
$\mu$, separately for the pseudoscalar and pseudovector pion-quark
coupling. All possible contributions are summed, coming from pion loops as well as
from gluon loops. 

\begin{figure}
        \centering
        \includegraphics[width= 6cm]{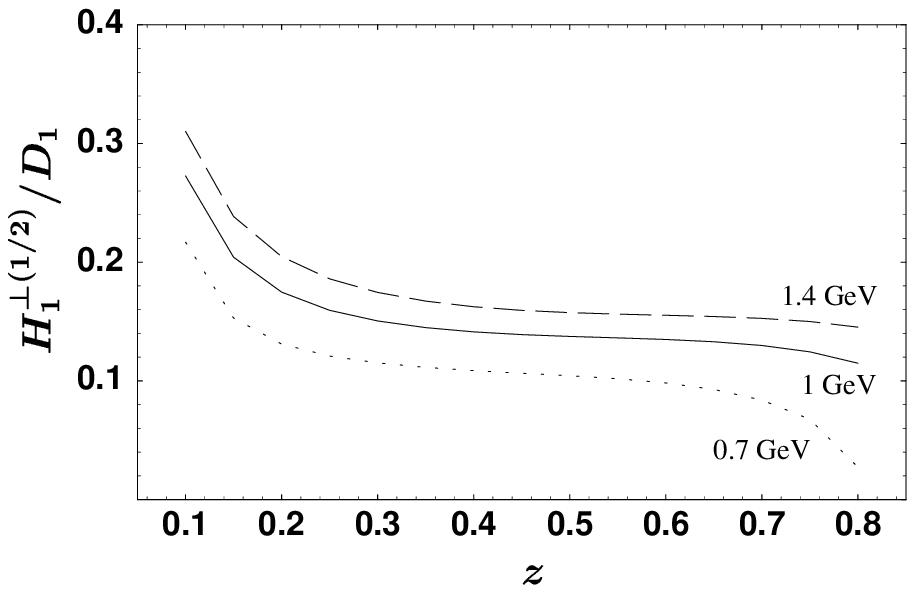}
        \hfil
        \includegraphics[width= 6cm]{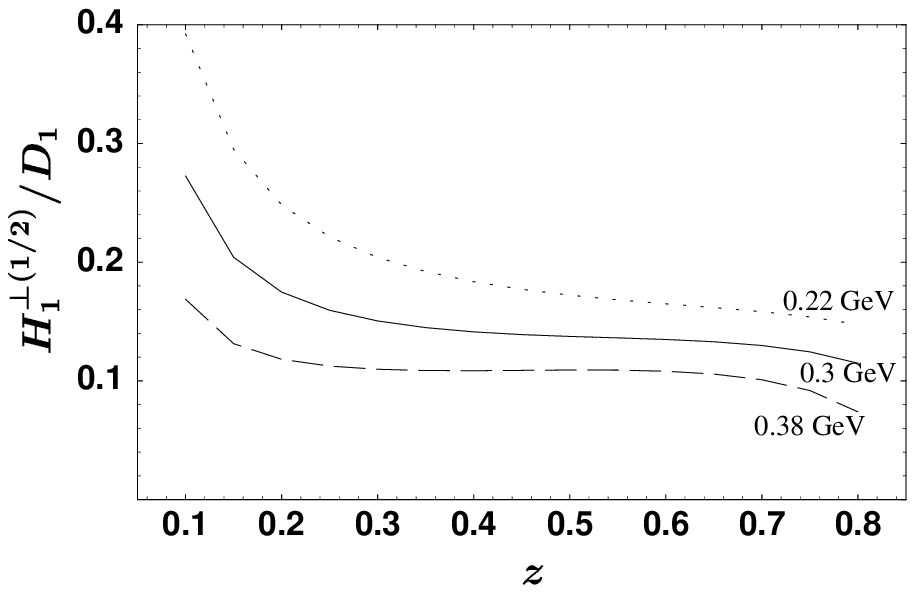}
        \caption{Estimate of $H_1^{\perp (1/2)}/D_1$ in a model with
          pseudoscalar pion-quark interaction, including pion and
          gluon single-loop corrections. Left panel: 
          dependence on the parameter $\mu$ (for $m=0.3$ GeV). Right panel:
          dependence on the parameter  $m$ (for $\mu=1$ GeV).} 
        \label{f:coll_ps}
        \end{figure}

Fig.~\ref{f:coll_ps} shows the result for the pseudoscalar coupling model of
Sec.~\ref{s:ps}. The left panel shows the dependence on the cutoff parameter
$\mu$ and the right panel the dependence on the constituent quark mass $m$. 
In this model 
the 
dominant part of the Collins function arises 
from gluon interactions, in particular from
the gauge-link box diagram, as can be seen by comparing the solid lines in
Fig.~\ref{f:coll_ps} with that in Fig.~\ref{f:coll_ps_g}. 
In general, this model gives rise to a ratio $H_1^{\perp (1/2)}/D_1$ of about
10-20\%, with a flat dependence on the variable $z$ in the range 
$0.2\leq z \leq 0.8$.

\begin{figure}
        \centering
        \includegraphics[width= 6cm]{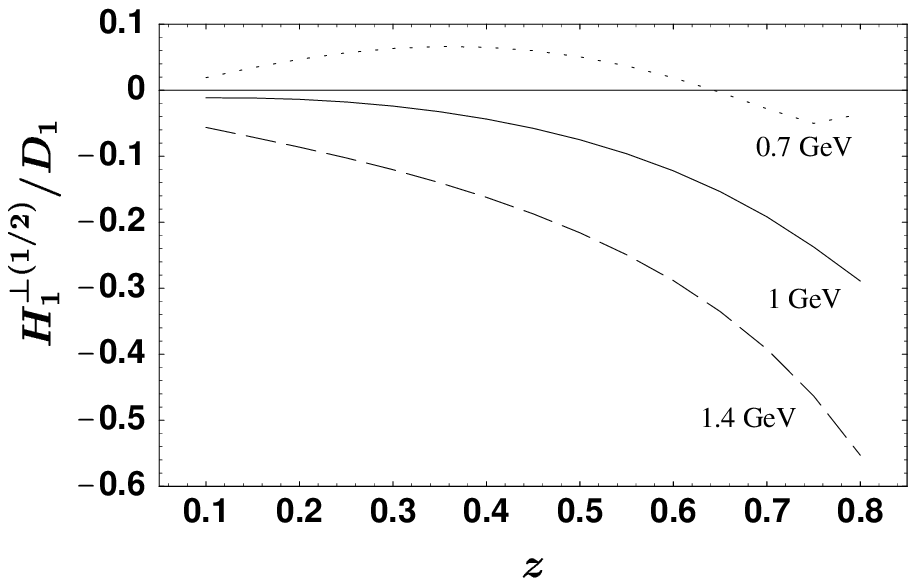}
        \hfil
        \includegraphics[width= 6cm]{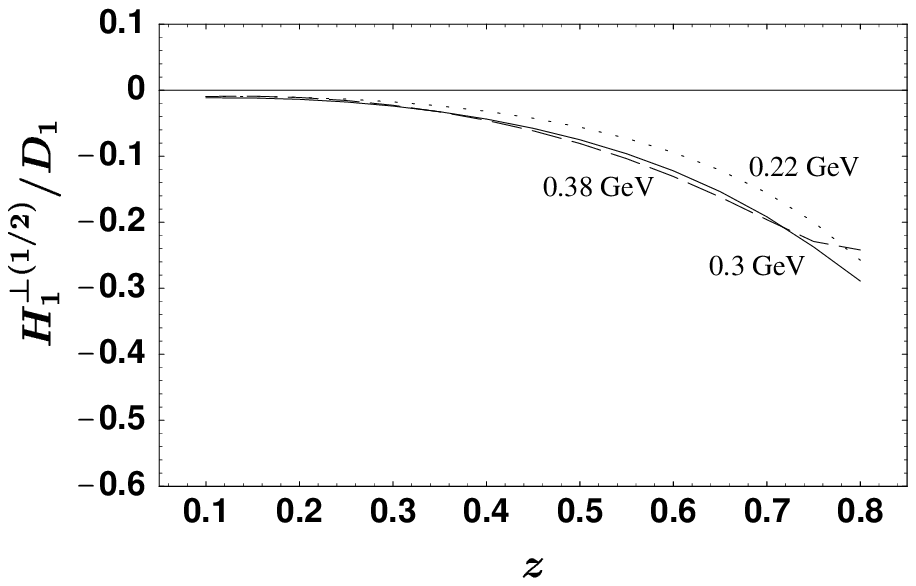}
        \caption{Estimate of $H_1^{\perp (1/2)}/D_1$ in a model with
          pseudovector pion-quark interaction, including pion and
          gluon single-loop corrections. Left panel: 
          dependence on the parameter $\mu$ (for $m=0.3$ GeV). Right panel:
          dependence on the parameter  $m$ (for $\mu=1$ GeV).}
        \label{f:coll_pv}
        \end{figure}

Fig.~\ref{f:coll_pv} shows the result for the pseudovector coupling model of
Sec.~\ref{s:psvec}. As before, the dependence on the cutoff parameter
$\mu$ and on the constituent quark mass $m$ is shown in the left and right
panel, respectively. In this case, for higher values of the parameter $\mu$
the contact-interaction contribution, diagram (c) of Fig.~\ref{f:pvpion}, is
the dominant one, making the ratio $H_1^{\perp (1/2)}/D_1$ negative and as big
as 30-50\%.
However, as the value of $\mu$ is reduced, the
size of the contact-interaction contribution decreases, while the size of the
contributions from the gluon loops increase. The value of the ratio
$H_1^{\perp (1/2)}/D_1$ can in this case become positive.

In order to allow a more direct comparison with experimental results, we
estimated the Collins transverse single-spin asymmetry for $\pi^+$ production
in semi-inclusive DIS. 
We use the definitions of the asymmetries and of the azimuthal angles
suggested in Ref.~\cite{Bacchetta:2004jz}. 
To shorten the notation, we introduce the Collins angle 
$\phi \equiv \phi_S+\phi_h$. We compute the
quantities~\cite{Mulders:1995dh,Bacchetta:2002tk} 
\begin{align}
 2 \,\frac{\int \frac{\de x \de y}{x
y^2}\, (1-y)\,\sum_a e_a^2\, h_1^a (x)\,H_1^{\perp (1/2)a}(z)}
{\int \frac{\de x \de y}{x y^2}\left(1-y+\frac{y^2}{2}\right)\sum_a e_a^2\, f_1^a
(x)\,D_1^a(z)} &\approx 2\,\left\langle \sin{\phi} \right\rangle_{UT}\, ,
\label{e:a_nw}
\\
2 \,\frac{\int \frac{\de x \de y}{x y^2}\,
         (1-y)\,z\,\sum_a e_a^2\, h_1^a (x)\,H_1^{\perp (1)a}(z)}
{\int \frac{\de x \de y}{x y^2}\,\left(1-y+\frac{y^2}{2}\right)\,\sum_a e_a^2\, f_1^a
(x)\,D_1^a(z)} &\approx 2\, \Big\langle \frac{|\vec{P}_{h\perp}|}{m_{\pi}}
\sin{\phi} 
\Big\rangle_{UT}\, , 
\label{e:a_w}
\end{align} 
where $a$ denotes the quark flavor.
The equivalence between the above formulae and experimental asymmetries is
approximate 
since power corrections and $\alpha_S$ corrections are not taken into account. 
Moreover, we assumed a full integration over the
outgoing hadron's transverse momentum and in Eq.~(\ref{e:a_nw}) 
no intrinsic quark 
transverse momentum
in the target is present.\footnote{For a discussion on how results are 
modified by introducing some initial transverse momentum see 
Ref.~\cite{Schweitzer:2003yr}.} 

For numerical calculations, we use the nonrelativistic assumption $h_1 = g_1$
and the simple parameterization
of $g_1$ and $f_1$ suggested in~\cite{Brodsky:1995kg}. 
In our model, disfavored fragmentation functions vanish, while this is not the
case in experiments. Therefore, we focus
only on $\pi^+$ production, where the impact of this limitation of the model 
should not be very relevant. We apply the kinematical cuts used in the HERMES
experimental paper~\cite{Airapetian:2004tw}.

        \begin{figure}
        \centering
        \includegraphics[width= 6cm]{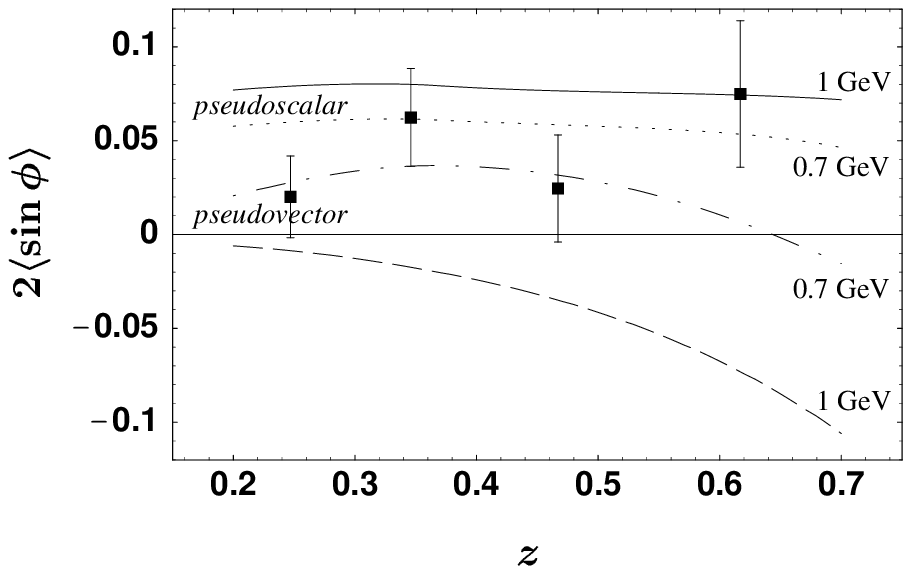}        
        \hfil
        \includegraphics[width= 6cm]{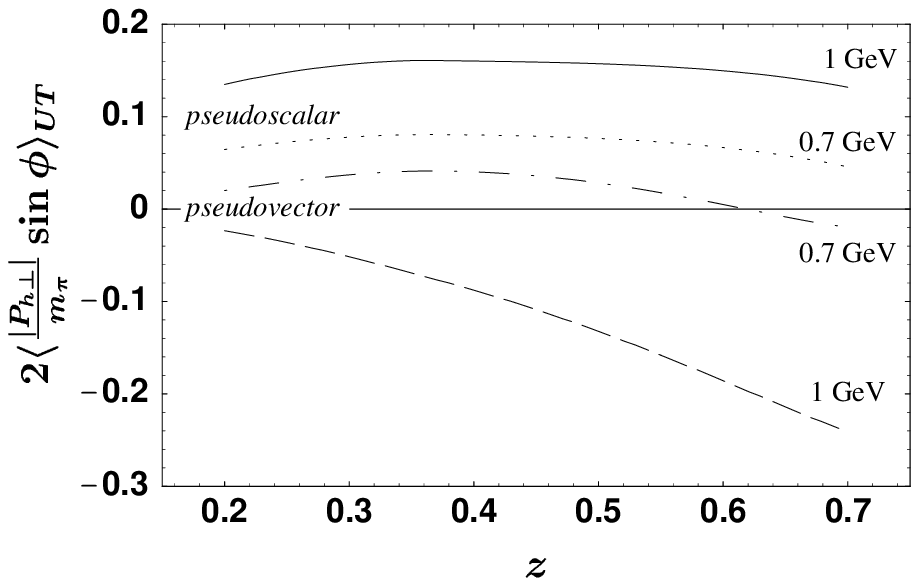} 
        \caption{Azimuthal transverse spin asymmetries
        $2 \langle\sin{\phi}\rangle_{UT}$ and 
        $2 \langle\sin{\phi}\,|\vec{P}_{h \perp}|/m_{\pi}\rangle_{UT}$
        for $\pi^+$ production at HERMES, 
        assuming $h_1 = g_1$ (see text), for both version of the
        pion-quark coupling and for two different values of the
        parameter $\mu$. Data points for the unweighted asymmetry are from \protect{\cite{Airapetian:2004tw}}.} 
        \label{f:az}
        \end{figure}

The results for the asymmetries are plotted in Fig.~\ref{f:az}. Published
HERMES data~\cite{Airapetian:2004tw} are also shown.  
We want to stress once more some caveats to
be taken into account when interpreting these results: 1 -- the models
could be modified by the introduction of form factors, 2 -- no calculation of
the real part of the loop corrections has been made, and disfavored
fragmentation functions vanish in this approach, 3 -- it is possible that the
measured asymmetry receives contributions not included in the
standard leading-order approach. 
Given the size of the experimental errors (only statistical errors are
included), it is difficult to discriminate between the models. Present data 
exclude the pseudovector coupling with a high value of
the parameter $\mu$. 
In general we observe that, making a reasonable choice of parameters,  we
 are not able to describe the data in the framework of the discussed models
 by using pionic degrees of freedom only.

\section{Conclusions}
\label{s:conc}

In this paper we reviewed four model calculations of the Collins
function for the fragmentation of a
quark into a pion,
which differ in the type of pion-quark coupling and in the type of one-loop
corrections they consider: 
pseudoscalar
pion-quark coupling with pion loops~\cite{Bacchetta:2001di} and with
gluon loops~\cite{Gamberg:2003eg}; pseudovector pion-quark coupling with
pion loops~\cite{Bacchetta:2002tk} and with gluon
loops~\cite{Bacchetta:2003xn}.
Even if these models have been already discussed in the
literature, we felt the need of presenting revised calculations in order to
fix some errors and discuss some details that were not explicitly addressed in
the past publications.
 
For all models, we discussed the shape and parameter dependence of the
unpolarized fragmentation function $D_1$. The agreement with typical
parameterizations of this function is not good for the pseudoscalar
coupling. For the pseudovector coupling, the comparison 
is more encouraging. 
The inclusion of
form factors in the pion-quark coupling can however change these results.

In all the above-mentioned approaches  
disfavored fragmentation functions
vanish, while data suggest that the disfavored Collins function could be as
big as the favored and have an opposite sign. In principle, disfavored
fragmentation functions could be calculated in the framework of the 
above-mentioned
models 
by considering diagrams with the emission
of two pions, one of which goes unobserved.

We computed the Collins function following the definition suggested in the 
``Trento
conventions''~\cite{Bacchetta:2004jz}. We found an overall sign error in
the published calculations of the Collins function in
Refs.~\cite{Bacchetta:2001di,Bacchetta:2002tk,Bacchetta:2003xn}. Moreover, 
we checked that
no contribution to the Collins function comes from the pole in the eikonal
propagator, in agreement with the general analysis of
Ref.~\cite{Collins:2004nx}. As a consequence, we pointed out that the
calculation of the Collins function in Ref.~\cite{Gamberg:2003eg} is in our
opinion wrong.

For one specific choice of the
parameters of the models, we analyzed separately the different
diagrams giving rise to a nonzero Collins function and we discussed the size
of their contribution to the final result. It turns out that in three out of
four cases the contributions of two diagrams roughly cancel each other,
so that the Collins function is driven dominantly by one diagram only. This
observation holds for different choices of the parameters, too. However, in
the case of the pseudovector coupling with gluon loops no dominant diagram can
be identified. 

The different signs obtained for the Collins function, and in particular for
the contributions of each diagram separately, lead us to the conclusion that
it is not possible to foresee the sign of the Collins
function a priori.
The fact that more than one
diagram with a different structure contributes to the Collins function makes
it also difficult, if not impossible, to interpret the resulting effect in
terms of 
simple attractive/repulsive interactions.

We studied the parameter
dependence of the ratio $H_1^{\perp (1/2)}/D_1$ for the two different choices
of the pion-quark coupling, summing together pion and gluon loop
corrections. Finally, we estimated the 
single-spin asymmetry for $\pi^+$ production
in semi-inclusive DIS off transversely polarized protons and compared it with
available experimental data~\cite{Airapetian:2004tw}.
The conclusions that can be drawn from the numerical results are somewhat
limited. It appears that 
data cannot be described by using pionic degrees of freedom only.
It seems also that data can exclude the 
pseudovector model with a high value of
the parameter $\mu$. All other versions of the models are compatible with the 
data.

\begin{acknowledgments}
The work of A.~B.\ has been
supported by the Alexander von Humboldt Foundation, the work of A.~M.\ by
the Sofia Kovalevskaya
Programme of the Alexander von Humboldt Foundation.
\end{acknowledgments}

\bibliographystyle{apsrev}
\bibliography{mybiblio}

\end{document}